\documentclass[aps,prd,reprint,groupedaddress]{revtex4-2}
\input epsf.sty
\usepackage{graphicx,amsmath,amssymb}

\def\lromn#1{\uppercase\expandafter{\romannumeral#1}}

\begin{document}


\title{
Axion cosmology in the presence of non-trivial Nambu-Goldstone
modes
}



\author{M. Yoshimura}
\affiliation{Research Institute for Interdisciplinary Science,
Okayama University \\
Tsushima-naka 3-1-1 Kita-ku Okayama
700-8530 Japan}


\date{\today}

\begin{abstract}
Axion cosmology is reexamined taking into account effect of kinetic pseudo Nambu-Goldstone modes,
with its importance recently pointed out.
When Peccei-Quinn (PQ) symmetry is broken by a chiral U(1) singlet,
it is found that the effect of kinetic Nambu-Goldstone mode makes the axion dark matter
untenable.
When PQ symmetry is extended and is broken by two singlets, we find axion cosmology to work,
but there are several differences from the axion cosmology studied in the literature.
The differences are (1) ordinary type of dark matter scaling as $1/{\rm cosmic \; scale\;
factor}^3$ arising from a modulus field  and not from the usual angular field,
(2) mass of the dark matter quantum in the ultralight range, $(10^{-32} \sim 10^{-14})\,$eV,
 (3) emergence of dark energy with the
present density of order (a few meV)$^4$ consistent with observations,
 (4) presence of a long range spin-dependent force,
(5) slow-roll inflation after PQ symmetry breaking when
conformal coupling to gravity is introduced.

\end{abstract}


\maketitle

\section{Introduction}

The invisible axion  \cite{axion 2}, \cite{axion 1} has been introduced 
to recover the  status of appealing solution to the problem of strong CP violation
\cite{pq and axion}, when the original axion  at electroweak scale  has
 been  excluded experimentally.
Fortunately it was later found that 
 the invisible axion may become one of the leading candidates for dark matter 
of our universe \cite{axion cosmology}, \cite{axiverse}.
The clue for this development is broken chiral  Peccei-Quinn (PQ) symmetry 
which generates pseudo Nambu-Goldstone (NG) field by QCD effect \cite{pq and axion}.
The symmetry breaking scale $f_a$ may be taken at will, but it is constrained by 
overclosure of the cosmic energy density and
 by  new cooling mechanism in stellar evolution.
If the PQ breaking scale $f_a$ is in the range of order
$10^{12}$ GeV, it was argued that the axion may become the major component of dark matter
close to the closure density \cite{axion cosmology}.

It was recently pointed out in another context \cite{my21}
that kinetic Nambu-Goldstone modes, usually
not considered in the mean field approach, give rise to a centrifugal repulsion contributing to
effective potential, thereby changing the potential minimum to determine the symmetry breaking scale.
When this idea is applied to a class of scalar-tensor gravity of cosmology, it 
is found \cite{my21} that the mechanism provides a simultaneous realization of 
dark energy and dark matter.

It is readily recognized that the same effect of kinetic NG modes
changes the standard scenario of axion cosmology.
The first part of the present work examines this problem, and indeed we find that
the axion cosmology fails in its original form.
We then present extension of PQ symmetry breaking scheme by
introducing two PQ singlets that enlarge the symmetry of scalar fields.
This extension saves the problem of the conventional axion cosmology, but
it brings in several new aspects to axion cosmology;
most notably, emergent dark energy along with a new form of dark matter,
with its quantum mass  in the ultralight range less than
$O(10^{-14})$ eV.
It has been advocated that ultralight dark matter  is appealing
from a variety of different motivations from ours, \cite{ultralight axion 1},  
\cite{ultralight 3}, \cite{axiverse}, \cite{ultralight axion 2}.

The present paper is organized as follows.
In Section \lromn2 we explain effect of kinetic Nambu-Goldstone modes and
show how this effect influences the axion cosmology discussed in the literature.
In Section \lromn3 Peceei-Quinn symmetry is extended for
a richer structure of NG modes in order to evade
the difficulty of conventional axion cosmology.
Section \lromn4 is the main part of this work that presents
axion cosmology in the extended model, stressing a number of
different consequences from the conventional result.
In Section \lromn5 we mention that introduction
of conformal coupling to gravity realizes a slow-roll inflation
after PQ symmetry breaking.
The paper ends with a brief summary.

We use the natural unit of $\hbar = c = 1$ 
throughout the present work  unless otherwise stated.

\section
{\bf Kinetic Nambu-Goldstone modes and problem of axion cosmology}

\subsection
{\bf Nambu-Goldstone modes}

Let us ignore QCD effect for simplicity and concentrate on NG modes,
in particular their kinetic contributions.
PQ chiral symmetry is realized in the simplest invisible axion model by
a complex singlet $\phi$ of appropriate U(1) charge, $\phi = \chi e^{i\theta}$ with
modulus and angular fields, $\chi\,, \theta,$ both taken real.
Potential $V(\phi) = V(\chi)$ is a function of the modulus field $\chi$ and is
independent of the angular field $\theta$.
Axion field is identified as $\chi \theta$, becoming $f_a \theta$
after spontaneous symmetry breaking where
$f_a$ is the field value at PQ symmetry breaking.
Kinetic terms, however, include $\theta$ and are written as
\begin{eqnarray}
&&
\frac{(\partial{\chi})^2}{2} + \frac{\chi^2 (\partial \theta)^2}{2}
\,.
\end{eqnarray}

Treating two fields, $\chi\,, \theta$, independently, the field equation for the
angular field becomes $\partial (\sqrt{-g} \, \chi^2 \partial \theta) = 0$,
 where a positive $-g$ is the metric determinant. This equation is integrated to give
$\sqrt{-g}\,\chi^2 \partial \theta = c_{\theta}$ (integration constant independent of spacetime position).
This constant is a quantum number of the angular momentum operator
in the abstract two dimensional field space $(\Re \phi\,, \Im \phi )$.
The generated centrifugal force may be incorporated in an effective potential as
$c_{\theta}^2/(-g\cdot 2\chi^2)$ \cite{my21}.
The effective potential $V_{\rm eff}$ has the form,
\begin{eqnarray}
&&
V_{\rm eff}(\chi) = \frac{c_{\theta}^2}{-g\cdot 2 \chi^2} + V(\chi)
\,.
\end{eqnarray}
Field location $\chi$ of potential minimum is shifted due to the presence of
centrifugal repulsive potential $\propto c_{\theta}$.
In cosmology this shift depends on the cosmic scale factor $R(t)$ since $-g = R^6(t)$
in the Robertson-Walker flat spacetime \cite{cosmology}.

\subsection
{\bf Field and Einstein equations}

Let us consider axion cosmology in which the potential is given by
$V(\chi\,, \theta) = \lambda \chi^2 (\chi^2 - 2 f_a^2)/4 - m_a^2 \chi^2 \cos \theta $.
A constant in the potential is fixed by  fine-tuning  the cosmological constant
such that $V(0 \,, \theta) = 0$.
We took the QCD potential $\propto \cos \theta$ using the dilute instanton approximation, but
the important points are the potential periodicity  in $\theta$ variable
and the proportionality to a small $m_a^2$, and not detailed form of the QCD potential.
This potential has a minimum at $\theta = 2\pi \times $ an integer
and $\chi = f_a$ giving the PQ symmetry breaking scale.
Field and Einstein equations for spatially homogeneous modes read as
\begin{eqnarray}
&&
\ddot{\chi} + 3 \frac{\dot{R}}{R} \dot{\chi} 
= \frac{c_{\theta}^2}{R^6 \chi^3} 
- \left( \lambda(\chi^2 - f_a^2) - 2 m_a^2 \cos\theta \right) \chi
\,,
\label {axion modulus field eq}
\\ &&
\ddot{\theta} + 3 \frac{\dot{R}}{R} \dot{\theta} + 2 \frac{\dot{\chi}}{\chi} \dot{\theta}
=  - m_a^2 \sin \theta
\,,
\label {axion angular field eq}
\\ &&
(\frac{\dot{R}}{R})^2 = \frac{1}{6 M_{\rm P}^2} \left( \frac{\dot{\chi}^2}{2}
+ \frac{ c_{\theta}^2}{2 R^6 \chi^2} 
+ \frac{\lambda}{4} \chi^2 (\chi^2 - 2 f_a^2)
\right.
\nonumber \\ &&
\left.
 - m_a^2 \chi^2 \cos \theta
 + \rho_M
\right)
\,,
\end{eqnarray}
where $\rho_M$ is contribution to energy density of nearly massless and massive particles
in the standard model.
The dot in the present paper means time derivative.
These equations are valid when QCD effects are turned on.
$M_{\rm P} = 1/\sqrt{16\pi G_N} \sim 1.72 \times 10^{18}$ GeV is
the Planck energy in our convention.

Two major differences from axion cosmology in the literature \cite{axion cosmology}
are (1) $c_{\theta} \neq 0$, (2) $\chi$ allowed to vary away from the constant $f_a$
and presence of $\dot{\chi}/\chi$ term in eq.(\ref{axion angular field eq}).

\subsection
{\bf Time evolution of axion and modulus fields}

The potential minimum of $V_{\rm eff}$ adiabatically (namely, slowly varying with time) shifts
due to the centrifugal repulsion $c_{\theta}^2/(2 R^6 \chi^2)$,
with the modulus field $\chi$ following the equation,
\begin{eqnarray}
&&
\chi\left( \lambda (\chi^2 - f_a^2) - 2 m_a^2 \right) = \frac{c_{\theta}^2}{ R^6 \chi^3} 
\,.
\label {potential minimum 1}
\end{eqnarray}
We took $\theta = 2\pi \times $ an integer (minimum location of angular variable).
Without the centrifugal repulsion ($c_{\theta}=0 $ case) the minimum is determined by vanishing left hand side.
A salient feature of centrifugal repulsion (the case $c_{\theta} \neq 0$)
 is an enormous change of the field value at
potential minimum, since the scale factor $R$ varies by roughly of order $10^{12}$
at two epochs, QCD and the present.

\begin{figure*}[htbp]
 \begin{center}
 \epsfxsize=0.4\textwidth
 \centerline{\epsfbox{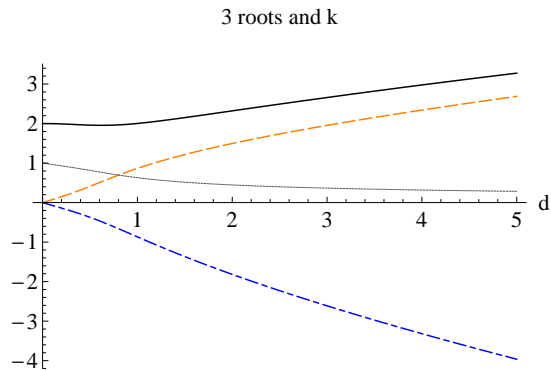}} \hspace*{\fill}\vspace*{1cm}
   \caption{
Three roots $a>b>c$ of algebraic equation appearing constant $R$ solution:
$a$ in solid black, $b$ in dashed orange, and $c$ in dash-dotted blue.
$k = \sqrt{(a-b)/(a-c)} $ factor in Jacobi solutions is shown in dotted black,
changing $\sim 1$ around $d=0$ to $\sim 0$ at $d \gg 1$.
}
   \label {three roots zero-e}
 \end{center} 
\end{figure*}

The most serious problem of conventional axion cosmology concerns the axion angular field
equation (\ref {axion angular field eq}), 
in particular, the presence of $\chi$ dependent term, $ 2\dot{\chi} \dot{\theta}/\chi$.
In order to quantify the effect of this term, let us first discuss time evolution
of the modulus field $\chi$ governed by (\ref{axion modulus field eq})
neglecting the small term proportional to the axion mass squared,
which is  justified at epochs prior to QCD.
Under the adiabatic approximation, there is a conservation law giving a constant energy $E$;
\begin{eqnarray}
&&
\frac{\dot{\chi}^2 }{2}  +
 \frac{ c_{\theta}^2}{2 R^6 \chi^2} + 
\frac{\lambda}{4} \chi^2  ( \chi^2-2  f_a^2)
= E
\,.
\end{eqnarray}
The $\chi$ equation is in a closed form for constant $R$, and is analytically integrated, giving a solution,
\begin{eqnarray}
&&
\int^{\chi_*^2}_{\chi ^2} dx\, \frac{1 }{\sqrt{ 2 (E - V_{\rm eff}(x)\,) } } = t
\,,
\\ &&
 V_{\rm eff}(\chi^2) =  \frac{ c_{\theta}^2}{2 R^6 \chi^2} + 
\frac{\lambda}{4} \chi^2  ( \chi^2-2  f_a^2)
\,.
\end{eqnarray}
The function
$\chi^2 (t)$ is given by Jacobi's elliptic function $ {\rm sn}(z; k)$ as
\begin{eqnarray}
&&
\hspace*{-0.5cm}
\chi^2 ( t)
= 
f_a^2  \left(a - (a-b) \, {\rm sn }^{2} (\frac{\sqrt{a-c}}{2} f_a t ; \sqrt{\frac{a-b}{a-c }})
\right)
\,,
\label {jacobi sols 1}
\end{eqnarray}
where parameters $a>b>c$ are three real roots of the third order algebraic equation,
\begin{eqnarray}
&&
x^2 ( x -2)  + d^2 = (x- a) (x-b) (x-c)
\,,
\label {3rd algebraic eq}
\\ &&
x = \frac{\chi^2}{f_a^2}
\,, \hspace{0.5cm}
d^2 = \frac{1}{ 2 R^6 } \frac{c_{\theta}^2}{f_a^6} 
\,.
\end{eqnarray}
We illustrate three roots $(a,b,c)$ and $k$ parameter of the elliptic function in Fig(\ref{three roots zero-e}).

The Jacobi elliptic function ${\rm sn}(z; k)$ has double periods in the complex $z$ plane,
and the relevant time period is $2/(f_a \sqrt{a-c}) $ times
\begin{eqnarray}
&&
K(k) = \int_0^{\pi/2 } \frac{d\varphi}{\sqrt{1- k^2 \sin^2 \varphi}} 
\,,
\end{eqnarray} 
with  slowly increasing $\sqrt{a-c} \rightarrow 1.54\, d^{1/3} $ as $d\rightarrow \infty$,
and always larger than 2,  as seen in Fig(\ref{three roots zero-e}).
Since it is natural to expect $d^2 \ll  O(1)$ at the present cosmological epoch,
the time period remains of order $1/f_a$, giving a rapid oscillation of the modulus $\chi$ field.

It would be instructive to show two limiting cases of Jacobi solutions;
$d \rightarrow 0^+$ and $d \rightarrow \infty $ limits.
In the $d \rightarrow 0^+$ limit, $\sqrt{a-b}\sim \sqrt{2}$ and $k \rightarrow 1^-$, which gives
\begin{eqnarray}
&&
\chi^2 \rightarrow f_a^2 \left( a - (a-b)\, {\rm tanh}^2 ( \frac{\sqrt{a-b}}{2} f_a t)
\right)
\,.
\end{eqnarray}
This is a delta-function with its peak value $a$  over a background $b$.
For a finite, but small $d$, the function is a periodic series of delta-functions.
On the other hand, in the $d \rightarrow \infty $ limit, $\sqrt{a-b}\sim 1.0\, d^{1/6} $ and 
$k \sim 0.65\, d^{-1/6} $.
The Jacobi solution is simplified to a sinusoidal function,
\begin{eqnarray}
&&
\chi^2 \rightarrow f_a^2 \left( a - (a-b) \sin^2  ( \frac{\sqrt{a-b}}{2} f_a t) \right)
\,.
\end{eqnarray}
Approaches to these limiting functions are not fast, but 
these formulas may help to understand the nature of solutions.

\begin{figure*}[htbp]
 \begin{center}
 \epsfxsize=0.4\textwidth
 \centerline{\epsfbox{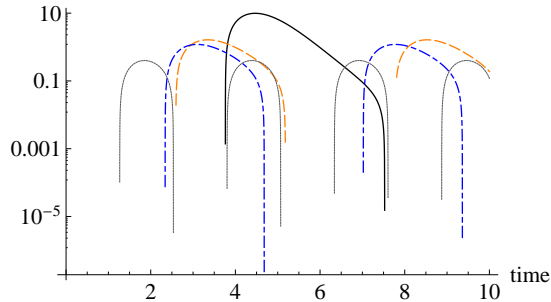}} \hspace*{\fill}\vspace*{1cm}
   \caption{
Time variation of 
acceleration force $2 \dot{\chi} c_{\theta}/\chi^3$ in unit of $f_a^2$ for axion angular field $\theta \sim a/f_a$:
0.1 in solid black, $\sqrt{16/27}$ in dashed orange, $1$ in dash-dotted blue,
and $5 $ in dotted black.
Time $t$ is scaled by the unit $2/(\sqrt{a-c}\, f_a)$ depending on $d$.
The time period average can never be in the required axion mass range of
$m_a \approx 10^{-27} f_a$.
}
   \label {accelerated rotation of axion angle}
 \end{center} 
\end{figure*}

With these in mind we analyze the axion angular evolution.
In the small axion mass limit it is sufficient to use the approximate equation,
\begin{eqnarray}
&&
\ddot{\theta} + 3 \frac{\dot{R}}{R} \dot{\theta} +  2 \frac{\dot{\chi}}{\chi^3} \frac{c_{\theta}}{R^3}
=  - m_a^2 \sin \theta
\,.
\end{eqnarray}
This is a damped pendulum equation with accelerating force given by the third term
$\propto c_{\theta}$ in the left hand side.
Conditions for damped oscillation necessary for dark matter interpretation 
are $m_a > 3 \dot{R}/{R}$ (for oscillation start) and $2 |c_{\theta} \dot{\chi}/\chi^3| < m_a^2  $ 
(condition to suppress continued pendulum rotation).
The second condition is difficult to satisfy, because $|c_{\theta} /\chi^3| \approx \sqrt{2}\,d$
at present and as numerically illustrated in Fig(\ref{accelerated rotation of axion angle}), 
$2 |c_{\theta} \dot{\chi}/\chi^3|  $ can never be in the appropriate axion squared mass range,
namely $< 1\, {\rm MeV}^2$ for its stability against $e^{\pm}$ pair annihilation.
The conclusion is robust despite that the adopted adiabatic approximation is not exact, and
damped oscillation never occurs when kinetic pseudo Nambu-Goldstone
modes are incorporated in the axion cosmology.
The result holds even if $\chi$ mode decays.

\section
{\bf Extended PQ symmetry breaking}

Since PQ mechanism is attractive as a solution to the strong CP problem,
we would like to explore extension of the original model \cite{axion 2}.
A possible direction towards extension would be to attribute the potential shift to another
field different from the axion.
For this purpose we introduce a second spinless field, this time
with no PQ-charge.

The simplest scheme toward improvements is
to introduce two complex spinless fields, $\phi_a = \phi_1 + i \phi_2$ and $\phi_d = \phi_3 + i \phi_4$
with $\phi_i\,, i = 1, \cdots 4$ real fields,
and identify PQ U(1) symmetry transformation as a subgroup of O(4) among four $\phi_i$:
$\phi_a \rightarrow e^{i \alpha X } \phi_a\,, \phi_d \rightarrow \phi_d$.
PQ charge $X$ is defined as in \cite{axion 2}.
$\phi_d$ does not transform under PQ symmetry transformation,
hence no direct coupling to two Higgs doublets, $H_u\,,H_d$.
But $\phi_d$ field has a non-trivial coupling to $\phi_a$ field in the  interaction  lagrangian,
$|\phi_a|^2 |\phi_d|^2$ arising O(4)-symmetric $( |\phi_a|^2 +  |\phi_d|^2)^2$.
We impose O(4) symmetry for the potential $V(\phi_i )$.
To quartic orders, we choose
\begin{eqnarray}
&&
V(\chi ) = \frac{\lambda_{\phi}}{4} \chi^2 (\chi^2 -2  f_a^2)
\,, 
\\ &&
\phi_a = \chi_a e^{i \theta_a}
\,, \hspace{0.3cm}
\phi_d = \chi_d e^{i \theta_d}
\,, \hspace{0.3cm}
\chi^2 = \chi_a^2 + \chi_d^2
\,.
\end{eqnarray}

The coupling potential to the Higgs sector
$V_{\phi H}=V_{\phi H}(H_u, H_d, \phi_a )$  is assumed to have
a partial O(4) symmetry transforming under U(1) PQ transformation, which involves $\phi_a$ singlet alone.
Thus, the direct $\phi_d$ coupling term 
$\phi_d^2  H_u^i \epsilon_{ij} H_d^j$ with Higgs doublets is forbidden by PQ symmetry,
while important $\phi_a^2  H_u^i \epsilon_{ij} H_d^j$ is allowed.
See below, however, on  induced $\phi_d$ coupling to Higgs doublets.
The concrete form of potential $V_{\phi H} $ is given in \cite{axion 2}.
Effect of the coupling $V_{\phi H} $ becomes important later when 
we discuss decay of heavy quantum.
For the time being we may ignore this coupling.

Pseudo NG kinetic terms may be incorporated in the an effective 
potential using a few quadratic terms of
six angular momentum components, $L_{ij}\,, i \neq j\,, i, j = 1\sim 4$:
\begin{eqnarray}
&&
V_{\rm eff}(\chi_a, \chi_d)  = V_{\rm NG} + V(\chi )
\,, 
\\ &&
V_{\rm NG} = 
\frac{\sum_{(ij)} L_{ij}^2}{2 R^6 \chi^2} + \frac{L_{12}^2}{ 2R^6 \chi_a^2} + \frac{L_{34}^2}{ 2R^6 \chi_d^2}
\,.
\end{eqnarray}
This particular pattern of symmetry breaking was chosen for our purpose.
The remaining symmetry is rotation among $(3,4)$ and $(1,2)$ components,
with the latter being broken by QCD effect.
The rest of kinetic terms is $\dot{\chi_a}^2/2 + \dot{\chi_d}^2/2$.
We denote quantum numbers of angular momentum components by
$c_a\,, c_d\,, c_{\psi}$ for three of $L_{12}^2\,, L_{34}^2\,, \sum_{(ij)} L_{ij}^2 $.

\section
{\bf Cosmology of extended axion model}

\subsection
{\bf Preliminary}

Field equations for $\chi_i, i = a,d$ and the Einstein equation are
\begin{eqnarray}
&&
\ddot{\chi_a} + \frac{3\dot{R}}{R} \dot{\chi_a} 
=
- \partial_{\chi_a} V_{\rm eff} (\chi_a, \chi_d, \theta_a) 
\,,
\\ &&
\ddot{\chi_d} + \frac{3\dot{R}}{R} \dot{\chi_d} 
=
- \partial_{\chi_d} V_{\rm eff} (\chi_a, \chi_d, \theta_a) 
\,,
\\ &&
(\frac{\dot{R}}{R})^2 = \frac{1}{6M_{\rm P}^2} \left(
\frac{\dot{\chi_a}^2}{2} + \frac{\dot{\chi_d}^2}{2} + V_{\rm eff} (\chi_a, \chi_d, \theta_a) 
\right)
\,,
\\ &&
V_{\rm eff} (\chi_a, \chi_d, \theta_a)  = \frac{c_a^2}{2 R^6 \chi_a^2} + \frac{c_d^2}{2 R^6 \chi_d^2}
+ \frac{c_{\psi}^2}{ 2 R^6 \chi^2}
\nonumber \\ &&
+ \frac{\lambda_{\phi}}{4} \chi^2 ( \chi^2 - 2 f_a^2) 
- m_a^2 \chi_a^2 \cos \theta_a 
\,.
\end{eqnarray}
These field equations are for spatially homogeneous modes.
For spatially inhomogeneous modes $\chi_i(\vec{x}, t)\,, i = a,d$ there are additional
derivative terms $- \vec{\nabla}^2 \chi_i /R^3$, which
are separately discussed later.
QCD term, $ - m_a^2 \chi_a^2 \cos \theta_a$,
 is proportional to a small squared mass  $m_a^2$, and approximate this
vanishing for some parts of subsequent discussion.
Angular fields obey
$\dot{\theta}_d = c_d/(R^3 \chi_d^2)\,, \dot{\psi}=c_{\psi}/(R^3 \chi^2) $,
while the equation for the axion angular field $\theta_a$
is more complicated and shall be discussed later.

We focus on cosmic time evolution at epochs when ordinary radiation and matter particle
contributions $\rho_m$ are negligible compared to dark fields.
The method we resort in these regions is  the adiabatic approximation, namely we assume
that the scale factor $R$ changes slowly, and approximate it as a constant
for some finite time region.
Potential minimum in this extended model is then given by two equations,
 $\partial_{\chi_a} V_{\rm eff} =
\partial_{\chi_d} V_{\rm eff} = 0$, and reads as
\begin{eqnarray}
&&
\chi_a \left( \lambda_{\phi} ( \chi^2 - f_a^2) - 2 m_a^2 \right) = 
\frac{1}{R^6} \left( \frac{c_a^2}{\chi_a^3} + \frac{c_{\psi}^2}{\chi^4} \chi_a \right)
\,,
\\ &&
\chi_d \lambda_{\phi} ( \chi^2 - f_a^2)  = \frac{1}{R^6} \left( 
\frac{c_d^2}{\chi_d^3} + \frac{c_{\psi}^2}{\chi^4} \chi_d \right)
\,.
\end{eqnarray}
We included the major QCD term, $- 2m_a^2$.
Dividing each of these by $\chi_i\,, i=a,d$ and
subtracting the second from the first or adding both, one derives
\begin{eqnarray}
&&
- 2 m_a^2 = \frac{1}{R^6} \left( \frac{c_a^2}{\chi_a^4}
- \frac{c_d^2}{\chi_d^4}  \right)
\,,
\\ &&
\hspace*{-0.3cm}
2 \left( \lambda_{\phi} ( \chi^2 - f_a^2) -  m_a^2 \right) = 
\frac{1}{R^6} \left( \frac{c_a^2}{\chi_a^4}
+  \frac{c_d^2}{\chi_d^4}   + 2 \frac{c_{\psi}^2}{\chi^4} \right)
\,.
\label {sum eq}
\end{eqnarray}
Thus, the quantity to be fine-tuned to the small axion mass 
is the difference $c_d^2/\chi_a^4 - c_a^2/\chi_d^4  $, 
and the total sum, the right hand side of eq.(\ref{sum eq}),  should be large of order,
$ \lambda_{\phi} ( \chi^2 - f_a^2) $.
In $m_a \rightarrow 0$ limit the global bifurcation into two modes is given by
\begin{eqnarray}
&&
\chi_i^2 = \frac{c_i}{c_a + c_d} \chi^2
\,, \hspace{0.3cm} i = a\,, d
\,, 
\label {chi bifurcation}
\\ &&
{\rm with}
\hspace{0.3cm}
\lambda_{\phi} \chi^4 ( \chi^2 - f_a^2) = \frac{c_{\psi}^2 + ( c_a + c_d)^2}{R^6}
\,.
\label {potential extrema}
\end{eqnarray}
For a finite $m_a$ there are additional terms, $\delta \chi_i^2 \propto m_a^2$.
One  needs $c_d \gg c_a$ for $ \chi_d^2 \gg \chi_a^2$.
The limit $c_a \rightarrow 0$ seems to give back the original axion cosmology.
But there are a number of differences, for instance $\chi_a^2 \ll \chi_d^2$.
We shall explain these as we proceed.
We denote extrema thus found by $(\chi_a^{(0)}(R) \,, \chi_d^{(0)}(R)\,)$.

Away from extrema,
two field equations for $\chi_a$ and $\chi_d$ are separately integrated when $m_a=0$, to give two
integration constants, $E_i\,, i = a,d$,
\begin{eqnarray}
&&
\hspace*{-0.3cm}
\frac{\dot{\chi_a}^2 }{2}  +
 \frac{ c_a^2}{2 R^6 \chi_a^2} +  \frac{c_{\psi}^2}{2 R^6  \chi^2 }
+ 
\frac{\lambda_{\phi}}{4} \chi^2  ( \chi^2-2  f_a^2)
= E_a
\,,
\\ &&
\hspace*{-0.3cm}
\frac{\dot{\chi_d}^2 }{2}  +
 \frac{ c_d^2}{2 R^6 \chi_d^2} +  \frac{c_{\psi}^2}{2 R^6  \chi^2 }
+ 
\frac{\lambda_{\phi}}{4} \chi^2  ( \chi^2-2  f_a^2)
= E_d
\,.
\end{eqnarray}

There are two ways to analyze this $(\chi_a, \chi_d)$ system:
(1) expansion of the effective potential $V_{\rm eff}$ around extrema,
(2) potential dominance approximation.
The second method is most useful at latest epochs of evolution.

\subsection
{\bf Expansion around extrema}

We shall first discuss intermediate epoch of scalar dominance using the method (1).
Deviation from extrema $(\chi_a^{(0)}(R) \,, \chi_d^{(0)}(R)\,)$ are denoted
by $(\delta \chi_a\,, \delta \chi_d)$ in terms of which the conservation equation
is approximately
\begin{eqnarray}
&&
\frac{(\dot{\delta \chi_a})^2 }{2}  
+ \delta \chi_a \left( \frac{\partial^2 V_{\rm eff}}{\partial \chi_a^2 } \right)_0 \delta \chi_a
+ \delta \chi_a \left( \frac{\partial^2 V_{\rm eff}}{\partial \chi_a \partial \chi_d } \right)_0
 \delta \chi_d
= E_a
\,,
\nonumber \\ &&
\label {chi-a conservation}
\\ &&
\frac{(\dot{\delta \chi_d})^2 }{2}  
+ \delta \chi_d \left( \frac{\partial^2 V_{\rm eff}}{\partial \chi_d^2 } \right)_0
 \delta \chi_d
+ \delta \chi_d \left( \frac{\partial^2 V_{\rm eff}}{\partial \chi_a \partial \chi_d } \right)_0
 \delta \chi_a
= E_d
\,,
\nonumber \\ &&
\label {chi-d conservation}
\\ &&
\hspace*{-0.3cm}
\left( \frac{\partial^2 V_{\rm eff}}{\partial \chi_a^2 } \right)_0
= \frac{1}{R^6} \left( 3 \frac{c_a^2}{\chi_a^4}  
+   \frac{(c_a+c_d)^2}{\chi^4} \right) + 4 (\lambda_{\phi} + \frac{c_{\psi}^2}{\chi^6} ) \chi_a^2 
\,,
\\ &&
\hspace*{-0.3cm}
\left( \frac{\partial^2 V_{\rm eff}}{\partial \chi_d^2 } \right)_0
= \frac{1}{R^6} \left( 3 \frac{c_d^2}{\chi_d^4}  
+   \frac{(c_a+c_d)^2}{\chi^4} \right) + 4 (\lambda_{\phi} +  \frac{c_{\psi}^2}{\chi^6} ) \chi_d^2
\,,
\\ &&
\left( \frac{\partial^2 V_{\rm eff}}{\partial \chi_a \partial \chi_d } \right)_0 =
4 \chi_a \chi_d (\lambda_{\phi} + \frac{c_{\psi}^2}{\chi^6} )
\,.
\end{eqnarray}
All quantities in $()_0$ are taken at $\chi_i^{(0)}(R)$.

Simplification of these formulas is made under the hierarchical parameter relation,
$c_{\psi} \gg c_d \gg c_a$:
\begin{eqnarray}
&&
\left( \frac{\partial^2 V_{\rm eff}}{\partial \chi_d^2 } \right)_0 \approx
4 \lambda_{\phi} f_a^2 + 8 \frac{c_{\psi}^2}{R^6}
\,,
\\ &&
\left( \frac{\partial^2 V_{\rm eff}}{\partial \chi_a \partial \chi_d } \right)_0
\approx 4 \sqrt{\frac{c_a}{c_d}} ( \lambda_{\phi} +  \frac{c_{\psi}^2}{\chi_d^6} ) \chi_d^2
\,,
\\ &&
\left( \frac{\partial^2 V_{\rm eff}}{\partial \chi_a^2 } \right)_0
\approx
4 \frac{c_a}{c_d}  \lambda_{\phi}\chi_d^2
\,,
\end{eqnarray}
by keeping at least one term for each of elements
and using the simplified form of eq.(\ref{potential extrema}):
\begin{eqnarray}
&&
\lambda_{\phi}\chi_d^4 (\chi_d^2 - f_a^2 ) \simeq \frac{c_{\psi}^2}{R^6}
\,.
\label {chi-d value}
\end{eqnarray}
This corresponds to eq.(\ref{potential minimum 1}) in the usual one-singlet model.

\begin{figure*}[htbp]
 \begin{center}
 \epsfxsize=0.4\textwidth
 \centerline{\epsfbox{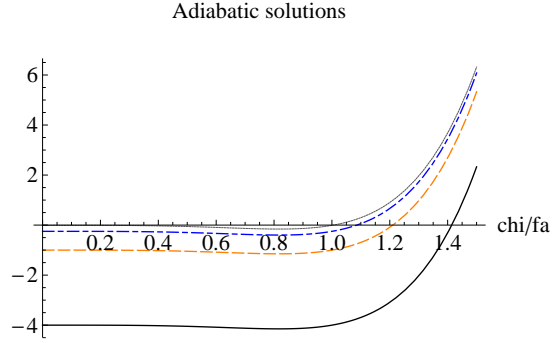}} \hspace*{\fill}\vspace*{1cm}
   \caption{
Solutions $\chi_d/f_a$ of eq.(\ref {chi-adiabatic sol}) for a few choices of $e$:
plot of  function $f(y) = y^4 ( y^2 - 1) - e^2$ for 
$e= 2$ in solid black, 1 in dashed orange, 0.5 in dash-dotted blue,
and 0.1 in dotted black.
}
   \label {adiabatic chi sols}
 \end{center} 
\end{figure*}

The conservation law for $\delta \chi_d$ is readily integrated, to give
oscillating solution,
\begin{eqnarray}
&&
\sqrt{\frac{ E_d}{4 ( \lambda_{\phi} f_a^2 + \frac{2 c_{\psi}^2}{R^6} )  }} 
\sin \left( 2 \sqrt{\lambda_{\phi} f_a^2 + \frac{2 c_{\psi}^2}{R^6} } \,(t- t_i)
\right)
\,.
\end{eqnarray}
Related $\delta \chi_a$ is solved directly from the equation involving second derivatives,
\begin{eqnarray}
&&
\ddot{\delta \chi_a} = - 4 \sqrt{\frac{c_a}{c_d}} \left( (\lambda_{\phi} + \frac{c_{\psi}^2}{\chi_d^6 })
\chi_d^2  \right)_0 \delta \chi_d
\,,
\end{eqnarray}
giving oscillatory behavior.
A possible linear growth term $\propto t$ is forbidden by the conservation equation,
eq.(\ref{chi-a conservation}),
and $\delta \chi_a$ oscillates with the same frequency as $\delta \chi_d$.
These indicate stability of adiabatic solutions $\chi_d^{(0)}\,, \chi_a^{(0)}$ with given
constant scale factor $R$.
The governing equation at the present epoch of
$R(t_0) =1$ may be written in terms of dimensionless quantities:
\begin{eqnarray}
&&
y^4 ( y^2 - 1) = e^2
\,, \hspace{0.5cm}
y = \frac{\chi_d}{f_a}
\,, \hspace{0.5cm}
e= \frac{c_{\psi}}{\sqrt{\lambda_{\phi}} f_a^3}
\,.
\label {chi-adiabatic sol}
\end{eqnarray}
Numerical solutions of this algebraic equation are illustrated in Fig(\ref{adiabatic chi sols}).
For $e \leq 0.1$ the solution gives $\chi_d/f_a \sim 1$.
If $e$ is close to, or smaller than, unity, the adiabatic solution indicates that
$\chi_d$ approaches to $f_a$ at the present epoch.

\subsection
{\bf Masses of  heavy and light quanta}

We shall discuss this problem in general terms, going back to $(\delta \chi_d, \delta \chi_a) $
part of sub-matrix:
\begin{eqnarray}
&&
{\cal W} = -
\left(
\begin{array}{cc}
\partial_{\chi_a}^2 V_{\rm eff} &  \partial_{\chi_a}\partial_{\chi_d}V_{\rm eff}   \\ 
\partial_{\chi_a}\partial_{\chi_d}V_{\rm eff}  &  \partial_{\chi_d}^2 V_{\rm eff} 
\end{array}
\right)_{\rm background}
\,.
\end{eqnarray}
This separation of two modes from angular modes is allowed in $m_a \rightarrow 0$ limit.
The effective potential is a sum of three functions,
$V_a(\chi_a^2) + V_d(\chi_d^2) + V_0(\chi_a^2 + \chi_d^2)$, hence
\begin{eqnarray}
&&
\partial_{\chi_i}^2 V_{\rm eff} = 4( V_i')'  \chi_i^2 + 2 V_i' + 4( V_0')'  \chi_i^2 + 2 V_0' 
\,, 
\\ &&
\partial_{\chi_a}\partial_{\chi_d}V_{\rm eff} = 4( V_0')'  \chi_a \chi_d
\,.
\end{eqnarray}
First derivative terms vanish at extrema, hence
$2\chi_i (V_i' +  V_0') = 0  $ gives $ V_i' +  V_0' = 0$ unless $\chi_i = 0$.
The secular equation of ${\rm det}(z + {\cal W}) = 0$ results in
\begin{eqnarray}
&&
z^2 - 4 \left( ( V_0')'  \chi^2 + ( V_a')'  \chi_a^2 + ( V_d')'  \chi_d^2
\right) z 
\nonumber \\ &&
+ 16 \chi_a^2 \chi_d^2 \left( ( V_a')' ( V_d')' +
( V_0')' (( V_a')' + ( V_d')' ) \right) =0
\,.
\end{eqnarray}
To go further, we must remember the hierarchical parameter relation $c_{\psi} \gg c_d \gg c_a$, which states that
$z^0$ order terms are sub-dominant.
It reduces the secular equation to an approximate one;
\begin{eqnarray}
&&
z^2 - 4 \left( ( V_0')'  \chi^2 + ( V_d')'  \chi_d^2
\right) z  =0
\,,
\end{eqnarray}
to leading and the next leading orders.
This result demonstrates that the first eigen-mode gives a heavy particle
a mass $2 \sqrt{( V_0')'  } = \sqrt{2 \lambda_{\phi}} f_a$,
while the second  has  zero mass, meaning actually a small mass much less than $f_a$.
More precisely, the light particle has 
a squared mass,
\begin{eqnarray}
&&
z_2 \simeq 4 \frac{ \chi_a^2 \chi_d^2 \left( ( V_a')' ( V_d')' +
( V_0')' (( V_a')' + ( V_d')' ) \right)  }{ ( V_0')'  \chi^2 + ( V_a')'  \chi_a^2 + ( V_d')'  \chi_d^2}
\sim \frac{4 c_a c_d}{R^6 f_a^4}  
\,.
\label{light particle mass}
\nonumber \\ &&
\end{eqnarray}
Near the present epoch this value is $\approx 4 c_a c_d/f_a^4$.
We  derive later an upper bound on this mass in eq.(\ref{dark quantum mass}).

As we shall show in the next subsection, the heavy particle and the corresponding field mode
quickly decay.
This has an important consequence of evading the overclosure of cosmic
density caused by heavy dark matter particles.

\subsection
{\bf Induced $\chi_d$ coupling to standard model particles and
its consequences}

$\chi_d - \chi_a$ mixing via $\chi_a$ coupling to Higgs doublets $H_u, H_d$
is expected to induce an effective coupling of $\chi_d$ to standard model particles.
These are important because inhomogeneous mode of $\chi_d$
as discussed in the following subsection has a mass very large and
it may have a serious difficulty of over-closing the universe.
In another word, $\chi_d$ matter, both homogeneous and inhomogeneous, must decay sufficiently fast.

First, note that our extended model has coupling term $\lambda_{\phi} |\phi_d|^2 | \phi_a|^2/2$,
which gives a three-point
vertex $\lambda_{\phi} \chi_d \delta \chi_d (\delta \chi_a)^2 $
and particle mixing $4\lambda_{\phi} \chi_d \chi_a \delta \chi_d \delta \chi_a $.
U(1) PQ non-singlet $\phi_a$ has coupling to Higgs doublet,
in the form $\chi_a \delta \chi_a H_u^i \epsilon_{ij} H_d^j $ \cite{axion 2}.
We ignore other terms of the form, $\chi_a \delta \chi_a |H_i|^2\,, i =u, d  $,
for simplicity.
In these formulas $\chi_i$ are background fields and
$\delta \chi_i$ are their quantum fields.
Thus, there is an induced $\delta \chi_d$ coupling to Higgs doublets of the form,
\begin{eqnarray}
&&
g_a\, \delta \chi_d H_u^i \epsilon_{ij} H_d^j
\,, \hspace{0.5cm}
g_a =4 \lambda_{\phi}\sqrt{\frac{ c_a}{c_d }} f_a
\,.
\end{eqnarray}
This gives rise to $\delta \chi_d$ (of mass $M \sim 2\sqrt{\lambda_{\phi}}\, f_a$) 
decay into a Higgs pair, $H_u\, H_d$.
Its decay rate is given by the formula,
\begin{eqnarray}
&&
\gamma_d = \frac{\lambda_{\phi}}{16 \pi} \frac{c_a}{c_d} M
\,,
\end{eqnarray}
ignoring small Higgs boson masses $\ll M$.
Placing constraint that decay is completed prior to QCD epoch,
$\gamma_d \geq H(\Lambda_{\rm QCD}) \approx O(4) \Lambda_{\rm QCD}^2/M_{\rm P} $, gives
\begin{eqnarray}
&&
\frac{c_a}{c_d} \geq \frac{8 \pi}{ \lambda_{\phi}^{3/2}} O(4)
\frac{\Lambda_{\rm QCD}^2 }{M_{\rm P} f_a } 
\,.
\label {early chi-d decay condition}
\end{eqnarray}
This is a condition that can be satisfied readily if $f_a$ is
not too small, for instance, $f_a > 10^9$ GeV.

Finally, we consider induced coupling of the massless mode $\theta_d$
previously judged as decoupled from the rest.
The question is that this mode couples to standard particles via the modulus
$\chi_d$, inducing a pseudo-scalar long range force.
The following field equation for this angular mode
\begin{eqnarray}
&&
\ddot{\theta_d} + 3 \frac{\dot{R}}{R} \dot{\theta_d} + 2 \frac{\dot{\chi_d}}{\chi_d} \dot{\theta_d}
= 0
\,,
\end{eqnarray}
indicates its coupling to $\chi_d$ field, which in turn gives an
induced coupling to Higgs doublets via $\chi_d-\chi_a$ mixing.
Detailed form of this induced coupling and its magnitude is left to future work.

\subsection
{\bf Potential dominated epoch and dark energy}

So far we relied on the adiabatic approximation.
We now turn to another approximation which focuses on time evolution
at latest times near the present.
When kinetic energy $\dot{\chi_a}^2/2 + \dot{\chi_d}^2/2$ is much less than
the potential energy $V_{\rm eff}$, one can approximate the exact
broken conservation equation,
\begin{eqnarray}
&&
\frac{d}{dt} \left( \frac{1}{2} \sum_i \frac{\dot{\chi}_i^2}{2} + V_{\rm eff} 
\right) = - 3 \frac{\dot{R}}{R} \sum_i \frac{\dot{\chi}_i^2}{2}
\,,
\end{eqnarray}
by  replacing $ \sum_i \dot{\chi}_i^2/2$ in the right hand side by 
$E -V_ {\rm eff}\,, E = E_a + E_d$ and dropping time variation of kinetic terms in the left hand side.
This leads to an approximate differential equation for the effective potential,
when it is supplemented by the Einstein equation,
\begin{eqnarray}
&&
\frac{d}{dt} V_{\rm eff} = - 3 \frac{1}{\sqrt{6} M_{\rm P}} \sqrt{V_{\rm eff}} 2 (E - V_{\rm eff})
\,.
\end{eqnarray}
This equation is readily integrated, to give for $t \geq t_i $
\begin{eqnarray}
&&
\sqrt{\frac{V_{\rm eff} (t) }{E}} =  \tanh \left( A_i - \sqrt{\frac{3 }{2 }} \frac{\sqrt{E} }{M_{\rm P} } ( t- t_i) \right)
\,, \hspace{0.3cm}
\\ &&
R(t) = R(i) \cosh^{-1/3}  \left( A_i - \sqrt{\frac{3 }{2 }} \frac{\sqrt{E} }{M_{\rm P} } ( t- t_i) \right)
\,,
\\ &&
A_i = {\rm arctanh} \sqrt{\frac{V_{\rm eff} (t_i) }{E}}
\,.
\end{eqnarray}
$R(i)$ is determined by the present value of scale factor $R(t_0) = 1$.

This is the solution relevant to the accelerating phase of late universe.
The Hubble constant at present $\dot{R}/R = H_0$ may be used to relate 
the dark energy density $E$ to other parameters.
By calculating
\begin{eqnarray}
&&
H_0 = \left(\frac{d}{dt} \ln R \right)_{t=t_0} =\frac{ \sqrt{ E}}{ \sqrt{ 6} M_{\rm P} }\sqrt{\frac{V_{\rm eff}(t_0)}{E}}
\,, 
\end{eqnarray}
one derives
\begin{eqnarray}
&&
E^{1/4} \sim 6^{1/4} \left( 1 - R^{-6}(i)\right)^{-1/4} \, \sqrt{H_0 M_{\rm P}} 
\nonumber \\ &&
\sim 6^{1/4} \sqrt{H_0 M_{\rm P}} 
\,, \hspace{0.3cm}
\sqrt{H_0 M_{\rm P}}  \approx 2\, {\rm meV}
\,.
\label {dark energy density}
\end{eqnarray}
Thus, independent of detailed evolution of scale factor and field,
the present dark energy density is found of order (a few meV)$^4$ consistent with
observations.
Time scale of variation is given by
\begin{eqnarray}
&&
\sqrt{\frac{2}{3}} \frac{M_{\rm P}}{\sqrt{E}} = \left( 1 - R^{-6}(i) \right)^{ 1/2} \frac{1}{3H_0}
\approx \frac{1}{3H_0}
\,,
\end{eqnarray}
roughly of order the cosmic age.
This should be compared to the oscillation period $T_o$ discussed in Subsection B and C,
which gave 
\begin{eqnarray}
&&
T_o =  \pi \frac{f_a^2 R^3}{  \sqrt{c_a c_d}}
\,.
\end{eqnarray}
Comparison of two formulas places a constraint,
\begin{eqnarray}
&&
 \frac{ \sqrt{c_a c_d}}{f_a^2}> 3\pi H_0
\,, \hspace{0.3cm}
H_0 \approx 10^{-33} {\rm eV}
\,.
\label {lower mass bound}
\end{eqnarray}

The dark energy density of the same order  has also been
derived by using a related, but somewhat different approach
 in the scalar-tensor gravity \cite{my21}.
Results of two approaches are similar.

The dark energy density $E$ given here may in principle include both 
of spatially homogeneous $\chi_d$ and $\chi_a$ modes,
but as discussed above, the heavy $\chi_d$ mode quickly decays if the inequality
(\ref{early chi-d decay condition}) is obeyed.
Thus, the dark energy consists of stable light $\chi_a$ mode.
Its field value $\chi_a^2$ is calculated by using the relation
$E = V_{\rm eff}(\chi_a)$ (effective potential containing $\chi_a$ mode alone).
Under the inequality constraint below, the present value of $\chi_a \equiv \chi_0$
satisfies the equation,
\begin{eqnarray}
&&
(\frac{c_a}{\chi_0^3})^2 = (\frac{m_a}{\chi_0})^2 + \lambda_{\phi}
\,,
\label {chi-a constraint 1}
\\ &&
(\frac{m_a}{\chi_0})^2 + \frac{3}{4} \lambda_{\phi} < 6  \frac{(H_0 m_{\rm P})^2 }{\chi_0^4 }
\,.
\label {chi-a constraint 2}
\end{eqnarray}
Since $\sqrt{H_0 M_{\rm P}}$ is of order a few meV, solutions in the same energy range is most natural,
giving both $\chi_0\,, (c_a)^{1/3}$ of order meV, and $m_a \leq $ a few meV
under the assumption, $\lambda_{\phi}$ of order unity.
The interesting possibility of $m_a$ much smaller than $O(\mu {\rm eV})$ is not
excluded.

We now discuss field variation $\dot{\chi}_a$ which becomes important to discussion
of axion dark matter in the next subsection.
The best way to derive a relation between $\dot{\chi}_a/\chi_a $
and the Hubble rate $\dot{R}/R$ is to use the adiabatic relation at
late times, $V_{\rm eff}(\chi_a; R) = E$ (a constant), to obtain
\begin{eqnarray}
&&
\frac{\dot{\chi}_a}{\chi_a} = \frac{ 6 c_a^2}{ R^6 \lambda_{\phi} \chi_a^6} \frac{\dot{R}}{R}
\,, \hspace{0.3cm}
\frac{6 \chi_a^2}{\lambda_{\phi} \chi_a^6} \sim 6 ( 1 + \frac{m_a^2 }{\lambda_{\phi} \chi_0^2 })
\,.
\label {chi0 and ca}
\end{eqnarray}
This implies that rate of $\chi_a$ variation is faster than the Hubble rate.
This peculiar situation of seemingly kinetic dominance
does not preclude an approximate constancy of dark energy density,
which shall be discussed after we introduce our final
model of conformal gravity in Section \lromn5.

\subsection
{\bf Spatially inhomogeneous modes as dark matter}

We next investigate time evolution of spatially inhomogeneous modes.
By taking homogeneous modes $\chi_{i,0}\,, \theta_{i,0} \,, i = a,d$
as a background, one analyzes inhomogeneous modes
by treating them in perturbative expansion or performing the linearized approximation.
Linearized modes are decomposed into modes of definite wave vectors;
 $\delta \chi_{i,k},\, \delta \theta_{i,k} \propto e^{i \vec{k}\cdot \vec{r} }$.
This is a valid procedure due to translational invariance of Robertson-Waler metric
and homogeneous solutions.
The field equations for amplitudes, $\delta \chi_{i,k},\, \delta \theta_{i,k}\,, i = a,d$, are
\begin{eqnarray}
&&
\hspace*{-0.5cm}
\left(
\frac{d^2 }{dt ^2} +  \frac{3}{R} \frac{d R}{dt} \frac{d }{dt } 
+ \frac{\vec{k}^2  }{ R^2} \right) 
\left(
\begin{array}{c}
\delta \chi_{a,k}   \\
\delta \theta_{a,k}   \\
\delta \chi_{d,k}   \\
\delta \theta_{d,k}  
\end{array}
\right)
= {\cal W} 
\left(
\begin{array}{c}
\delta \chi_{a,k}   \\
\delta \theta_{a,k}   \\
\delta \chi_{d,k}   \\
\delta \theta_{d,k}  
\end{array}
\right)
\,,
\\ &&
\hspace*{1cm}
{\cal W} = -
\nonumber \\ &&
\left(
\begin{array}{cccc}
\partial_{\chi_a}^2 V_{\rm eff} &  \partial_{\chi_a}\partial_{\theta_a}V_{\rm eff}  
&\partial_{\chi_a}\partial_{\chi_d}V_{\rm eff}   & 0 \\ 
\partial_{\chi_a}\partial_{\theta_a}V_{\rm eff}/\chi_a^2  & \partial_{\theta_a}^2 V_{\rm eff}/\chi_a^2  
&0 & 0 \\
\partial_{\chi_a}\partial_{\chi_d}V_{\rm eff}  & 0 & \partial_{\chi_d}^2 V_{\rm eff}  & 0\\
0&0&0&0
\end{array}
\right)_{{\rm background}}
\,.
\end{eqnarray}
The $\delta \theta_d$ mode appears to decouple from the rest, giving
a long range massless field.
As discussed in Subsection D, there exists, however, an induced coupling
of $\delta \theta_d$ to standard model particles via Higgs doublets,
leading to a long range spin-dependent force.

The rest of three components $(\delta \chi_{a,k}\,,\delta \theta_{a,k} \,, \delta \chi_{d,k}  ) $ 
may or may not be coupled,
 and this system is analyzed using $ 3\times 3$ sub-matrix $- {\cal W} $ in the form,
\begin{eqnarray}
&&
\hspace*{-0.5cm}
\left(
\begin{array}{ccc}
4 \sqrt{\frac{c_a}{c_d}} ( \lambda_{\phi} +  \frac{c_{\psi}^2}{f_a^6} ) f_a^2
\,, &  0 \,, 
& 4 \frac{c_a}{c_d}  \lambda_{\phi} f_a^2    \\ 
0 \,,& - m_a^2\,, 
&0  \\
4 \frac{c_a}{c_d}  \lambda_{\phi} f_a^2 \,,   & 0\,, & 
4 \lambda_{\phi} f_a^2 + 8 \frac{c_{\psi}^2 }{f_a^6}
\end{array}
\right)
\,.
\end{eqnarray}
The three component system is thus further decomposed into a single angular mode $\delta \theta_{a,k} $
and two-component coupled system  $(\delta \chi_{a,k} \,, \delta \chi_{d,k}  )$.

The inhomogeneous angular field $\delta \theta_{a,k} $ satisfies 
the differential equation,
\begin{eqnarray}
&&
\ddot{\delta \theta }_{a,k} + ( 3 \frac{\dot{R}}{R} + 2 \frac{\dot{\chi}_a}{\chi_a})\dot{\delta \theta}_{a,k}
= - ( m_a^2 +  \frac{\vec{k}^2  }{ R^2})\delta \theta_{a,k}
\,.
\end{eqnarray}
The presence of the third term $\propto \dot{\chi}_a/\chi_a$ is
characteristic of pseudo Nambu-Goldstone mode.
As discussed in the preceding subsection,
this term, with eq.(\ref{chi0 and ca}), damps the oscillation faster than the Hubble expansion
rate $3\dot{R}/R$.
The modified picture of dark axion field at latest times is as follows.
When  $2 \dot{\chi}_a/\chi_a  $ becomes smaller than $m_a$,
the field starts to oscillate with frequency $ \sqrt{m_a^2 + \vec{k}^2/R^2}$
and amplitude $\propto 1/R^{6}$. This inverse power $-6$ is subject to the
precise $\chi_a$ variation rate at late times, but it appears that the decrease rate
is faster than that of radiation.
Note, however, that there exists another candidate of cold dark matter
behaving like an ordinary one with its number decrease rate $\propto 1/R^3$,
as we shall see shortly.

The symmetry breaking bound $f_a < O(10^{13})$GeV 
derived in the literature \cite{axion cosmology} is
based on a few assumptions: the epoch of axion damped oscillation starting at $m_a = 3\dot{R}/R$
with the number density decrease $\propto 1/R^3$, 
and the current algebra estimate relating axion parameters to pion parameters,
$f_a m_a \approx f_{\pi} m_{\pi}$ \cite{b-tye},
\cite{coupling}.
Since the initial time of damped oscillation is changed at $m_a \sim 6\dot{R}/R$ and
the damping rate $\propto 1/R^6$ is faster in our extended model,
this quoted bound on $f_a$ is no longer applicable.

The two-component system $(\delta \chi_{a,k}, \delta \chi_{d,k}) $ consists of two particles moving 
with the momentum $\vec{k}/R$;
a decaying heavy particle and a light stable particle  forming
a candidate of dark matter.
The mass of light  particle predominantly made of $\chi_a$ mode is given by a time dependent 
function, eq.(\ref{light particle mass}), 
and its value $\sim 2 \sqrt{c_a c_d}/f_a^2 $ at the present epoch.
In Subsection E it was suggested that $c_a^{1/3}$ is of order, a few meV,
which implies that the mass $m_D$ of light particle is constrained by
\begin{eqnarray}
&&
m_D =
\frac{\sqrt{c_a c_d}}{f_a^2 } \ll O(10^{-14} {\rm eV}  )(\frac{10^{9} {\rm GeV}}{f_a} )^{1/2}
\,,
\label {dark quantum mass}
\end{eqnarray}
with $c_d \ll c_{\psi} \approx f_a^3$.
As a reference $f_a$  value we took astrophysics bound  by
stellar cooling derived in the standard axion cosmology \cite{astro constraint}.
This should be re-examined by using the  coupling scheme of new dark matter
different from the original axion model.
The lower bound, eq.(\ref{lower mass bound}), gives the inequality,
$m_D > 3\pi H_0 \approx 10^{-32} {\rm eV}$.

According eq.(\ref{light particle mass}), 
the light dark matter particle has its mass inversely proportional to 
the scale factor to cube, hence at nucleo-synthesis it is likely that
the mass is of order 100 GeV or slightly less, taking the Peccei-Quinn symmetry
breaking scale at $10^{12}$GeV.
There is no abundant particle close to thermal values at nucleo-synthesis from which
this massive particle can be produced.
The problem persists even at higher temperatures, because the heaviest
particle in standard particle physics is Higgs boson.
A likely production process is then radiative emission
from excited ion states in cosmic plasma after nucleo-synthesis:
\begin{eqnarray}
&&
{\rm He}^{+,*} + \gamma \rightarrow {\rm He}^+ + \chi
\,,
\end{eqnarray}
when the mass is below the ionization energy $\sim 20 $eV, and
at redshifts below $\sim 10^5$.
Produced energy spectrum is non-thermal, and there is no significant
interaction of produced dark matter particle with the rest
of cosmic constituents.
It is thus expected that a significant part of dark matter particles
is in the non-relativistic region.
Even if there exists a significant relativistic component,
this part loses its role in the energy budget compared to the cold component
due to its slower decrease rate.

Precise estimate of produced cold dark matter amount
requires detailed calculation of radiative emission rate,
which is beyond the scope of the present work.

An interesting possibility of detecting ultralight dark matter clouds surrounding black holes
by using gravitational wave emission has been pointed out \cite{axiverse}.
The method appears to work due to a recent announcement of the bound 
in the mass range $\sim 10^{-13}$ eV (frequency $20 \sim 610$ Hz)
for scalar dark matter \cite{sr-gw 1}.

\subsection
{\bf Nature of dark energy and dark matter and
comic microwave background anisotropy}

The equation-of-state factor $w$  characterizes
radiation, matter, and dark energy in cosmology based on general relativity.
$\Lambda$ CDM cosmology given by a mixture
of $w=0$ CDM and $w=-1$ dark energy has been
successful in explaining cosmic microwave background (CMB)
anisotropy assuming nearly scale invariant fluctuation
at recombination epoch.

In conformal gravity the role of $w$ factor is less obvious
due to a modified form of the energy-momentum conservation.
In the Einstein frame of conformal gravity they take different forms:
\begin{eqnarray}
&&
T_{\mu\nu}^{(E)} (\chi) = \frac{5}{F} \left(
\partial_{\mu} \chi \partial_{\nu} \chi
- \frac{1}{2} g_{\mu \nu}\partial_{\alpha}\chi \partial^{\alpha} \chi \right)
+ g_{\mu \nu} \frac{ V_{\phi}(\chi^2)}{F^2}
\,,
\nonumber \\ &&
\label {energy-momentum tensor in e-frame}
\\ &&
\dot{\rho_{\chi}}^{(E)}  = - ( 3 \frac{\dot{R} }{R} +  \frac{\partial{\chi} F}{F} \dot{\chi})
\, \frac{5}{ F(\chi)} \rho_{\chi}^{(E)} ( 1 + w_{\chi}^{(E)})
\,,
\label {energy conservation in e-frame}
\\ &&
\rho_{\chi}^{(E)} = 
\frac{5}{2F} \dot{\chi}^2 + \frac{5(\vec{\nabla}\chi)^2}{2F R^2} + \frac{V_{\phi}(\chi^2)}
{F^2}
\,,
\\ &&
w_{\chi}^{(E)}\rho_{\chi}^{(E)} = p_{\chi}^{(E)} = 
\frac{5}{2F} \left(
\dot{\chi}^2 - \frac{(\vec{\nabla}\chi)^2}{3 R^2}
\right) - \frac{V_{\phi}(\chi^2)}
{F^2}
\,.
\end{eqnarray}
Derivation of the energy-momentum
tensor in the Einstein frame, eq(\ref{energy-momentum tensor in e-frame}),
 is given for instance in \cite{my91}.
These relation holds separately for heavy and light inflaton, $\chi_d\,, \chi_a$.

With the presence of a large suppression due to 
the conformal factor $\propto 1/F(\chi)$
in the right-hand side of eq.(\ref{energy conservation in e-frame}),
the energy density $\rho_{\chi}^{(E)}$ can decrease towards a constant, 
irrespective of
$w_{\chi}^{(E)}$ values of order unity (but not necessarily equal to $-$ unity).
Indeed, one can derive, by neglecting complicated contribution
$\propto 3\dot{R}/(RF)$ in the right-hand side
(which actually gives a faster approach towards the constant behavior), 
$\rho_{\chi} = {\rm const.}\times \exp[ 5(1 + w_{\chi}^{(E)})/F]$ for a constant 
$w_{\chi}^{(E)}$ in the range $-1 \leq w_{\chi}^{(E)} \leq 1$.
This explains why the energy density stays nearly constant, although
there may be a non-trivial contribution from kinetic energy $\dot{\chi}^2/2$
in $\rho_{\chi}^{(E)}$.
Indeed, our model does not always give a stable value of $w_{\chi}^{(E)}=-1$
at some times of cosmic evolution.

Nevertheless, CMB anisotropy is sensitive to the pressure contribution
$p_{\chi}^{(E)}$ around the recombination epoch.
We shall therefore directly calculate the pressure and $w^{(E)}$
for the  dark energy and CDM proposed in preceding subsections.
Around the recombination epoch the  dark energy is made of
a light inflaton $\chi_a$, its homogeneous component , while CDM is made of 
inhomogeneous modes around this component $\delta \chi_a$.
The dark energy field $\chi_a$ is given by eq.(\ref{chi bifurcation}) and (\ref{chi-d value}):
 \begin{eqnarray}
&&
\chi_a^2 \sim \frac{c_a}{c_d} \chi_d^2 \sim 
\frac{c_a}{c_d}(\frac{c_{\psi}^2 }{ R^6 \lambda_{\phi}} )^{1/3 } =
\frac{c_a  c_{\psi}^{2/3 }}{ \lambda_{\phi}^{1/3 }c_d R^2 }
\,.
\end{eqnarray}
As argued around (\ref{chi-a constraint 1}) and (\ref{chi-a constraint 2}),
this quantity $\chi_a$ is of order  meV.
The potential term contributes to the dark energy density, which is calculated 
by balancing the Nambu-Goldstone
kinetic repulsion.
The result is  $3 \lambda_{\phi} \chi_a^4/4$.
The kinetic contribution given from (\ref{chi0 and ca}) is much smaller than this value.
Therefore,
\begin{eqnarray}
&&
\rho_{DE} \sim \lambda_{\phi} \,O({\rm meV})^4
\,,
\end{eqnarray}
and the equation-of-state factor $w_{DE} = -1$.

On the other hand,
CDM inflaton $\delta \chi_a$ has a mass $m_D$ given by (\ref{dark quantum mass}),
and  its Fourier-mode equation is given by
\begin{eqnarray}
&&
\ddot{\delta \chi_a} + 3 \frac{\dot{R}}{R} \dot {\delta \chi_a} 
+ \frac{q^2}{R^2}\delta \chi_a = - m_D^2 \delta \chi_a
\,.
\end{eqnarray}
CDM energy density $\rho_{CDM}$ varies with time according to
\begin{eqnarray}
&&
\frac{d}{dt} ( R^3 \rho_{CDM}) = \frac{1}{2} (\delta \chi_a)^2
\frac{d}{dt}
\left( R^3 ( m_D^2 + \frac{q^2}{ R^2})
\right)
\,.
\nonumber \\ &&
\end{eqnarray}
The right-hand side decreases  fast enough to prove the
CDM energy density decrease
$\propto 1/R^3$ and effectively $w_{CDM} = 0$.
Thus, CDM anisotropy spectrum is well described by
$\Lambda$ CDM model   in the present work as well.

\section
{\bf Other important issue of cosmology}

As discussed in \cite{my21},
the scalar-tensor gravity including conformal coupling realizes both inflation
and accelerating universe at late times.
This brings in a new aspect to axion cosmology when applied to
our extended O(4) model:
inflation after PQ-symmetry breaking.
We shall discuss this issue.

The conformal coupling of $\phi_d$ field to gravity is introduced 
in the Jordan-Brans-Dicke metric frame as
\cite{jordan}, \cite{brans-dicke}
\begin{eqnarray}
&&
\hspace*{-0.5cm}
{\cal L} =\sqrt{\overline{-g}} \left(
{\cal L}_{\phi g} + {\cal L}_{\rm EW}(H_u, H_d, \phi_a, \psi, \overline{g_{\mu\nu}})
+ {\cal L}_{\rm QCD}
\right)
\,, 
\\ &&
\hspace*{-0.5cm}
{\cal L}_{\phi g} =
- M_{\rm P}^2 F(\chi_d) \overline{R} 
+ \frac{1}{2} \sum_{i=a,d} \partial \phi_i^{\dagger} \partial \phi_i
- V_{\phi}(\chi_d^2 + \chi_a^2)
\,.
\\ &&
F( \chi_d) = 1 + \xi \frac{\chi_d^2}{f_a^2}
\,, \hspace{0.3cm}
\xi > 0
\,,
\end{eqnarray}
where $\overline{R}(\overline{g_{\mu\nu}} ) $ is the Ricci scalar curvature.
If $\xi = 0$, this model coincides with the  extended model in preceding sections.
In a sense this new model is a merger of O(2) scalar-tensor gravity \cite{my21} and
O(2) axion model into an enlarged O(4) symmetric scheme.
It is often convenient to Weyl rescale into the Einstein metric frame by using a new metric 
$g_{\mu\nu} = F(\chi_d) \overline{ g_{\mu\nu}}$:
\begin{eqnarray}
&&
\hspace*{0.5cm}
\sqrt{\overline{-g} }{\cal L}_{\phi g} 
\nonumber \\ &&
\hspace*{-0.5cm}
= \sqrt{-g} \left( - M_{\rm P}^2 R(g_{\mu\nu})
+ \frac{5}{2 F} \sum_{i=a,d} \partial \phi_i^{\dagger} \partial \phi_i - \frac{1}{F^2} V_{\phi}
\right)
\,.
\label {pq conformal coupling}
\end{eqnarray}

At inflationary epochs the $\phi_d$ field is dominant over $\phi_a$ field, hence
we shall drop contributions from the $\phi_a$ field.
Scalar kinetic term $\propto \partial \phi_d^{\dagger} \partial \phi_d $ 
is transformed to the standard form by a field re-definition that
introduces $\overline{\chi}_d$,
\begin{eqnarray}
&&
\hspace*{-0.3cm}
\overline{\chi}_d = \sqrt{5} \int^{\chi_d}_0 \frac{d u}{\sqrt{ F(u)} }
\,, \hspace{0.3cm}
\chi_d = \frac{f_a}{\sqrt{\xi}} \sinh (\sqrt{\frac{ \xi}{5}} \frac{\overline{\chi}_d}{f_a})
\,.
\end{eqnarray}
For small $\chi_d$, $\overline{\chi}_d  \simeq \sqrt{5} \chi_d $.
Using $ V_{\phi} = \lambda_{\phi} \chi_d^2 (\chi_d^2 - 2 f_a^2 )/4$,
one has
\begin{eqnarray}
&&
\frac{1}{F^2} V_{\phi} = \frac{\lambda_{\phi} f_a^4 }{4 \xi^2 }(1 - \frac{1}{F}) ( A - \frac{B}{F})
\,, 
\\ &&
A = 1
\,, \hspace{0.3cm}
B = 1 + 2 \xi
\,.
\end{eqnarray}
The positivity of potential $V(\chi_d)$ requires $A - \frac{B}{F} >0 $, or
\begin{eqnarray}
&&
1+ \xi \frac{\chi_d^2}{f_a^2} > \frac{B}{A} \; (= 1 + 2 \xi)
\,,
\end{eqnarray}
which is readily satisfied if $\chi_d > \sqrt{2} f_a $.

At earliest epochs of $\chi_d > O(f_a/\sqrt{\xi})$ 
the slow-roll inflation \cite{inflation models 1} may be realized as discussed in \cite{my21}.
The slow-roll conditions are imposed on the potential derivatives \cite{cosmology}
\begin{eqnarray}
&&
| \frac{V_{\rm eff}' }{V_{\rm eff} } | \ll \frac{1}{\sqrt{\xi }} \frac{f_a}{M_{\rm P}}
\,, \hspace{0.3cm}
| \frac{V_{\rm eff}'' }{V_{\rm eff} } | \ll \frac{3}{ 2\xi} ( \frac{f_a}{M_{\rm P}})^2
\,,
\\ &&
V_{\rm eff} (\eta) = \frac{ \lambda_{\phi} f_a^4}{4 \xi^2} \frac{1}{ (1 + \eta^2)^2} 
\nonumber \\ &&
\times \left(
A - \frac{A+B}{F} + \frac{B}{F^2} + \frac{2 \xi ^3 ( c_{\psi}^2 + c_d^2) }{\lambda_{\phi} R^6 f_a^6 }\frac{1}{\eta^2}
\right)
\,,
\end{eqnarray}
where the prime $'$ indicates field derivative with respect to 
the dimensionless variable $\eta = \sqrt{\xi }\chi_d/f_a (>0)$.
The two derivative conditions reduce to an identical inequality,
\begin{eqnarray}
&&
\chi_d \gg 2 M_{\rm P}
\,.
\end{eqnarray}
If $\chi_d$ at inflation is of order $f_a$, we derive $f_a \gg 2 M_{\rm P} $.
The slow-roll inflation ends around local minimum producing standard model particles
in thermal equilibrium.

Thus, inflation, dark energy, and dark matter can be
 a manifested pattern of bifurcated symmetry breaking with cosmological evolution,
as pointed out in \cite{my21}, but this time solving the strong CP problem at the same time.
Nevertheless, the fine-tuning of cosmological constant must be artificially introduced
as usual.

Skeptics might question why one has to consider a higher energy state of
non-vanishing centrifugal repulsion.
Inflation after PQ symmetry breaking gives an answer,
because our universe after inflation happens to be a fluctuation from the lowest
energy state, as in the chaotic inflationary scenario.
This view has a sort of anthropic flavor.

\section
{\bf Summary}

Kinetic Nambu-Goldstone modes have drastic effect to
axion cosmology: the potential minimum that determines the
field value of Peccei-Quinn symmetry breaking is shifted due to
time dependent centrifugal repulsion caused by NG modes.
This change makes the cosmology of original one-singlet model untenable,
but its extension to two singlet models saves axion cosmology
without losing the attractive feature as a solution to the
strong CP problem.
Nevertheless, axion cosmology based on the extended model 
is very different so that the major detection strategy of dark matter 
 focused on $\mu$eV range \cite{admx} should be reconsidered,
shedding more light on the ultralight mass range,
$(10^{-32} \sim 10^{-14})$ eV.

\vspace{0.5cm}
{\bf Note added in proof}

After submitting this work for publication,
we calculated the spectral index $n_s$ and the tensor-to-scalar ratio  $r$
of curvature fluctuation in an extended model that incorporates
a bare cosmological constant $\Lambda$ by changing $M_{\rm P}^2 F \bar{R}$
in eq.(\ref{pq conformal coupling}) to $M_{\rm P}^2 (F  \bar{R}+ 2 \Lambda)$.
The  fine-tuning condition of cosmological constant in the Einstein frame
is given by
\begin{eqnarray}
&&
2 \Lambda M_{\rm P}^2 + \lambda_{\phi} f_a^4 (\frac{f_a}{M_{\rm P}})^2 
\left( (\frac{f_a}{M_{\rm P}})^2 - 1
\right) = 0
\,,
\end{eqnarray}
without any problem to impose.

The spectral index $n_s$ and the tensor-to-scalar ratio  $r$
 are given in terms of inflaton potential derivatives, and
we calculated these, using formulas given  in \cite{kami-kov}.
Results of the above model  are as follows:
\begin{eqnarray}
&&
r = (2 \lambda_{\phi}  \xi)^2 \frac{ f_a^4 \chi_d^6 }{\Lambda^2 M_{\rm P}^6}
\,,
\\ &&
n_s - 1 + \frac{3}{8} r = 2 \lambda_{\phi} \xi \frac{f_a^2 \chi_d^2}{ \Lambda M_{\rm P}^2}
\,.
\end{eqnarray}
Formulas are valid approximately at large values of field $\chi_d$ during
the slow-roll inflation.
BICEP/Keck \cite{bicep/keck} and Planck \cite{planck} observations indicate
that $n_s - 1 \sim 0.035\,, r \leq 0.036$.
Our result is made consistent with these data by
properly arranging field magnitude $\chi_d$ during inflation
for given parameters $\Lambda, f_a$ and couplings $\lambda_{\phi}, \xi$.
The result is however model dependent, and if one takes
a quartic form for conformal function $F(\chi)$, 
results change, leaving more freedom.
Detailed analysis shall be published elsewhere.

This result on $n_s$ and $r$ differs from interesting
and related  conformal $\alpha$ attractor model of \cite{linde et al}.

\vspace{0.5cm}
\begin{acknowledgments}
This research was partially
 supported by Grant-in-Aid   21K03575   from the Japanese
 Ministry of Education, Culture, Sports, Science, and Technology.

\end{acknowledgments}

\end{document}